\documentclass[prd,superscriptaddress,showpacs,nofootinbib,amsmath,amssymb,aps,10pt]{revtex4}

\usepackage{bm}
\usepackage{amsfonts}
\usepackage{latexsym}
\usepackage[latin1]{inputenc}
\usepackage{graphicx}
\usepackage{amsmath}
\usepackage{palatino}
\usepackage{mathpazo}
\usepackage[outdir=./]{epstopdf}

\usepackage{booktabs}
\usepackage{dcolumn}

\def\jnl@style{\it}
\def\aaref@jnl#1{{\jnl@style#1}}

\def\aaref@jnl#1{{\jnl@style#1}}

\def\aj{\aaref@jnl{AJ}}                   
\def\apj{\aaref@jnl{ApJ}}                 
\def\apjl{\aaref@jnl{ApJ}}                
\def\apjs{\aaref@jnl{ApJS}}               
\def\apss{\aaref@jnl{Ap\&SS}}             
\def\aap{\aaref@jnl{A\&A}}                
\def\aapr{\aaref@jnl{A\&A~Rev.}}          
\def\aaps{\aaref@jnl{A\&AS}}              
\def\mnras{\aaref@jnl{Mon.~Not.~Roy.~Astron.~Soc.}}             
\def\prd{\aaref@jnl{Phys.~Rev.~D}}        
\def\prc{\aaref@jnl{Phys.~Rev.~C}}  
\def\prl{\aaref@jnl{Phys.~Rev.~Lett.}}    
\def\qjras{\aaref@jnl{QJRAS}}             
\def\skytel{\aaref@jnl{S\&T}}             
\def\ssr{\aaref@jnl{Space~Sci.~Rev.}}     
\def\zap{\aaref@jnl{ZAp}}                 
\def\nat{\aaref@jnl{Nature}}              
\def\aplett{\aaref@jnl{Astrophys.~Lett.}} 
\def\apspr{\aaref@jnl{Astrophys.~Space~Phys.~Res.}} 
\def\physrep{\aaref@jnl{Phys.~Rep.}}      
\def\physscr{\aaref@jnl{Phys.~Scr}}       
\def\commat{\aaref@jnl{Comm.~Math.~Phys.}}              
\def\science{\aaref@jnl{Science}}               
\def\cqg{\aaref@jnl{Classical Quant.~Grav.}}            
\def\jpcs{\aaref@jnl{JPCS}}                                     
\def\ijmpd{\aaref@jnl{Int.~J.~Mod.~Phys.~D}}                    
\def\grg{\aaref@jnl{Gen.~Relat.~Gravit.}}               
\def\rpp{\aaref@jnl{Rep.~Prog.~Phys.}}          
\def\npa{\aaref@jnl{Nucl.~Phys.~A}}        
\def\lrr{\aaref@jnl{Living Rev.~Rel.}}                   
\def\jcap{\aaref@jnl{J.~Cosmology Astropart.~Phys.}}    
\def\rmp{\aaref@jnl{Rev.~Mod.~Phys.}}   

\allowdisplaybreaks[1]

\begin{document}

\title{Rapidly rotating  neutron stars in $R$-squared gravity}
\author{Stoytcho S. Yazadjiev}
\email{yazad@phys.uni-sofia.bg}
\affiliation{Department
	of Theoretical Physics, Faculty of Physics, Sofia University, Sofia
	1164, Bulgaria}
\affiliation{Theoretical Astrophysics, Eberhard-Karls University
	of T\"ubingen, T\"ubingen 72076, Germany}

\author{Daniela D. Doneva}
\email{daniela.doneva@uni-tuebingen.de}
\affiliation{Theoretical Astrophysics, Eberhard-Karls University
	of T\"ubingen, T\"ubingen 72076, Germany}
\affiliation{INRNE - Bulgarian Academy of Sciences, 1784  Sofia, Bulgaria}

\author{Kostas~D.~Kokkotas}
\email{kostas.kokkotas@uni-tuebingen.de}
\affiliation{Theoretical Astrophysics, Eberhard-Karls University
	of T\"ubingen, T\"ubingen 72076, Germany}

\begin{abstract}
	$f(R)$ theories of gravity are one of the most popular alternative explanations for dark energy and therefore studying the possible astrophysical implications of these theories is an important task. In the present paper we make a substantial advance in this direction by considering rapidly rotating neutron stars in $R^2$ gravity. The results are obtained numerically and the method we use is non-perturbative and self-consistent. The neutron star properties, such as mass, radius and moment of inertia, are studied in detail and the results show that rotation magnifies the deviations from general relativity and the maximum mass and moment of inertia can reach very high values. This observation is similar to previous studies of rapidly rotating neutron stars in other alternative theories of gravity, such as the scalar-tensor theories, and it can potentially lead to strong astrophysical manifestations.
\end{abstract}


\maketitle

\section{Introduction}

The interest in studying the modifications of general relativity is growing significantly in the past few years. From one side this is connected with our aim to better understand the theory of gravity and to explore what are the possible deviations. But on the other hand there are observations which do not completely fit into the standard framework of Einstein's theory of gravity, such as the dark energy phenomena. Exactly this phenomena is the main motivation for the recent interest and advance of $f(R)$ theories of gravity which can offer an alternative and viable explanation for the dark energy.

Naturally $f(R)$ theories of gravity were mainly investigated in a cosmological context. But every theory of gravity has to be able to also pass various astrophysical tests. As far as black holes are concerned, no-hair theorems exist \cite{Heusler1995,Sotiriou2010,DeFelice2010a,Nojiri2011}, according to which the solutions in $f(R)$ theories are the same as the ones in general relativity (in the case of linear matter sources). But this is not the case with neutron stars, where the presence of matter can lead to significant deviations. On the other hand there is a variety of observations of compact stars that can be used to test both the weak field and the  strong field regime of different theories of gravity. That is why studying neutron stars in $f(R)$ theories and confronting them against the observations, is an inseparable part of the attempts to find a generalized theory of gravity that can naturally incorporate the dark energy phenomena.

Compact stars in  $f(R)$ theories were intensively studied in the past few years \cite{Cooney2010,Babichev2010,Arapoglu2011,Jaime2011,Santos2012,Orellana2013,Alavirad2013,Astashenok2013,Astashenok2014}. Due to the complexity of the field equations, the more detailed investigation of the problem and the confrontation with observations were done primarily using a perturbative approach (see e.g. \cite{Cooney2010,Arapoglu2011}). But it was later shown that studying the neutron stars in $f(R)$ theories of gravity perturbatively is misleading and one has to solve the full field equations self-consistently \cite{Yazadjiev2014,Staykov2014}. As a matter of fact the calculations can be simplified considerably if one uses the mathematical equivalence between the $f(R)$ gravity and a particular class of scalar-tensor theories with a nonzero potential for the scalar field.

The examination of nonrotating compact stars in $f(R)$ theories showed that constraints on the parameters can not be set via the present observations of the neutron star masses and radii alone, since the uncertainties in the nuclear matter equation of state can be bigger than the deviations from the pure general relativity \cite{Yazadjiev2014}. But the results in the case of slow rotation suggest a possible test of $f(R)$ gravity via the expected future observations of the neutron star moment of inertia \cite{Lattimer2005,Kramer2009}, that can impose tight constrains on the free parameters in the theory \cite{Staykov2014}. A natural extension of these results is to consider rapid rotations. An astrophysical motivation comes from the observations of millisecond pulsars and the models of supramassive neutron stars formed after a merger, where the rotational effects will be non-negligible. On the other hand it was shown in \cite{Doneva2013} that rotation can significantly increase the deviations from general relativity in the case of scalar-tensor theory. Taking into account that $f(R)$ theories are mathematically equivalent to them, we can expect a similar magnification of the differences, that can potentially lead to additional observations constraints on  $f(R)$ theories at astrophysical scales.

\section{Basic equations}
We will describe the analytical setup of the problem very briefly. The reader is referred to \cite{Yazadjiev2014,Staykov2014,Doneva2013} for a detailed discussion of the problem.

The action in $f(R)$ theories of gravity is given by
\begin{eqnarray}\label{A}
	S= \frac{1}{16\pi G} \int d^4x \sqrt{-g} f(R) + S_{\rm
		matter}(g_{\mu\nu}, \chi),
\end{eqnarray}
where $R$ is the Ricci scalar curvature with respect to the space-time metric $g_{\mu\nu}$, $S_{\rm matter}$ is the action of the matter, and the matter fields are collectively denoted by $\chi$. The inequalities $d^2f/dR^2\ge 0$ and $df/dR>0$ have to be imposed in order for the theory to be free of tachyonic instabilities and the appearance of ghosts. In the present paper we will be concentrated on the so-called $R^2$ theory of gravity for which
\begin{equation}
f(R) = R + aR^2,
\end{equation}
where $a$ is a parameter satisfying $a\ge 0$ in agreement with the above mentioned inequalities.

In our calculations we will use the fact that the $f(R)$ theories are mathematically equivalent to the Brans-Dicke theory with a parameter $\omega_{BD}=0$ and nonzero potential of the scalar field. The Jordan frame action is given by
\begin{eqnarray}
	S=\frac{1}{16\pi G} \int d^4x \sqrt{-g}\left[\Phi R - U(\Phi)\right]
	+ S_{\rm matter}(g_{\mu\nu}, \chi),
\end{eqnarray}
where the gravitational scalar $\Phi$  and the  potential $U(\Phi)$ are defined by $\Phi=\frac{df(R)}{dR}$ and
$U(\Phi)=R \frac{df}{dR} - f(R)$, respectively.  In the particular case of $R^2$ gravity, that we are considering, one can easily show that $\Phi=1 + 2aR$ and  the scalar-field potential is $U(\Phi)=\frac{1}{4a}(\Phi-1)^2$.

Even though the Jordan frame is the physical frame where the physical quantities are measured, in practice it is often more convenient to study the scalar-tensor theories in the so-called Einstein frame, where the metric  $g^{*}_{\mu\nu}$ is defined by the following conformal transformation
$g^{*}_{\mu\nu}=\Phi g_{\mu\nu}$. The action in Einstein frame can be written in the form

\begin{eqnarray}\label{EFA}
	S=\frac{1}{16\pi G} \int d^4x \sqrt{-g^{*}}\left[ R^{*} - 2
	g^{*\mu\nu}\partial_{\mu}\varphi \partial_{\nu}\varphi - V(\varphi)
	\right] + S_{\rm
		matter}(e^{-\frac{2}{\sqrt{3}}\varphi}g^{*}_{\mu\nu},\chi),
\end{eqnarray}
where $R^{*}$ is the Ricci scalar curvature with respect to the Einstein frame metric $g^{*}_{\mu\nu}$ and the new scalar field $\varphi$ is defined by $\varphi =\frac{\sqrt{3}}{2}\ln\Phi$. The Einstein frame potential $V(\varphi)$ is correspondingly $V(\varphi)=A^4(\varphi)U(\Phi(\varphi))$ where the Einstein frame coupling function $A(\varphi)$ is given by
\begin{equation}
A^2(\varphi)=\Phi^{-1}(\varphi)=e^{-\frac{2}{\sqrt{3}}\varphi}.
\end{equation}
For the particular case of $R^2$ gravity the scalar-field potential takes the form
\begin{equation}
V(\varphi)= \frac{1}{4a}
\left(1-e^{-\frac{2\varphi}{\sqrt{3}}}\right)^2.
\end{equation}

Due to the significant simplification of the field equations in the Einstein frame we will employee it\footnote{The Einstein frame quantities will be marked with an asterisk.}. Since this is not the physical frame, one has to transform the quantities, such as mass, radius and moment of inertia, back to the physical Jordan frame. A more detailed discussion  of the problem can be found in \cite{Yazadjiev2014,Staykov2014,Doneva2013}.

Since we will consider rotating neutron stars, the following general ansatz for the stationary and axisymmetric Einstein frame metric can be used
\begin{eqnarray}
	&&ds_{*}^2 = -e^{\gamma+\sigma} dt^2 + e^{\gamma-\sigma} r^2
	\sin^2\theta (d\phi - \omega dt)^2 + e^{2\alpha}(dr^2 + r^2
	d\theta^2),
\end{eqnarray}
where all the metric functions depend on $r$ and $\theta$ only. The dimensionally reduced field equations for the metric functions, the scalar field $\varphi$ and the equation for hydrostationary equilibrium can be found in the Apendix (see also \cite{Doneva2013}). The quantities, such as mass $M$ and  angular momentum $J$, are calculated using integrals throughout the volume of the star and the details are given in \cite{Doneva2013}. The moment of inertia of the star is given by the standard formula $I=J/\Omega$, where $\Omega$ is the angular frequency.

The field equations are solved using an extended version of the {\tt RNS} code \cite{Stergioulas95} where the modifications coming from the scalar-tensor theory are implemented \cite{Doneva2013}. But the code developed in  \cite{Doneva2013} is only for the case of zero scalar field potential. Here we extend it by taking into account the nontrivial scalar field potential $V(\varphi)$ which is directly connected to the choice of the function $f(R)$. Such an extension is nontrivial, because the introduction of a potential makes the scalar field equation very stiff. That is why one has to be very careful with the choice of the auxiliary parameters, such as the numerical accuracy and the relaxation factor, in order to be able to converge to the desired solutions. In addition to the tests performed in \cite{Doneva2013}, we have verified our code against the results obtained for static and slowly rotating neutron stars in $R^2$ gravity \cite{Yazadjiev2014,Staykov2014}.

In the next section, where the numerical results are presented, we shall use the dimensionless parameter $a\to a/R^2_{0}$  and the dimensionless moment of inertia $I\to I/M_{\odot}R^2_{0} $, where $M_{\odot}$ is the solar mass and $R_{0}$ is one half of the solar gravitational radius   $R_{0}=1.47664 \,{\rm km}$ (i.e. the solar mass in geometrical units).

\section{Numerical results}
In our studies we have chosen two realistic equations of state (EOS) that fulfill the current observational constraints on the neutron star masses and radii \cite{Lattimer12,Steiner2010,Ozel2013,Antoniadis13,Demorest10}. These are the APR4 \cite{AkmalPR} and  SLy4 EOS \cite{Douchin2001}. This choice is motivated also by the fact that both of them were used in our previous studies of static and slowly rotating neutron stars in $f(R)$ gravity \cite{Yazadjiev2014,Staykov2014}, that eases the comparison of the results.

\begin{figure}[]
	\centering
	\includegraphics[width=0.9\textwidth]{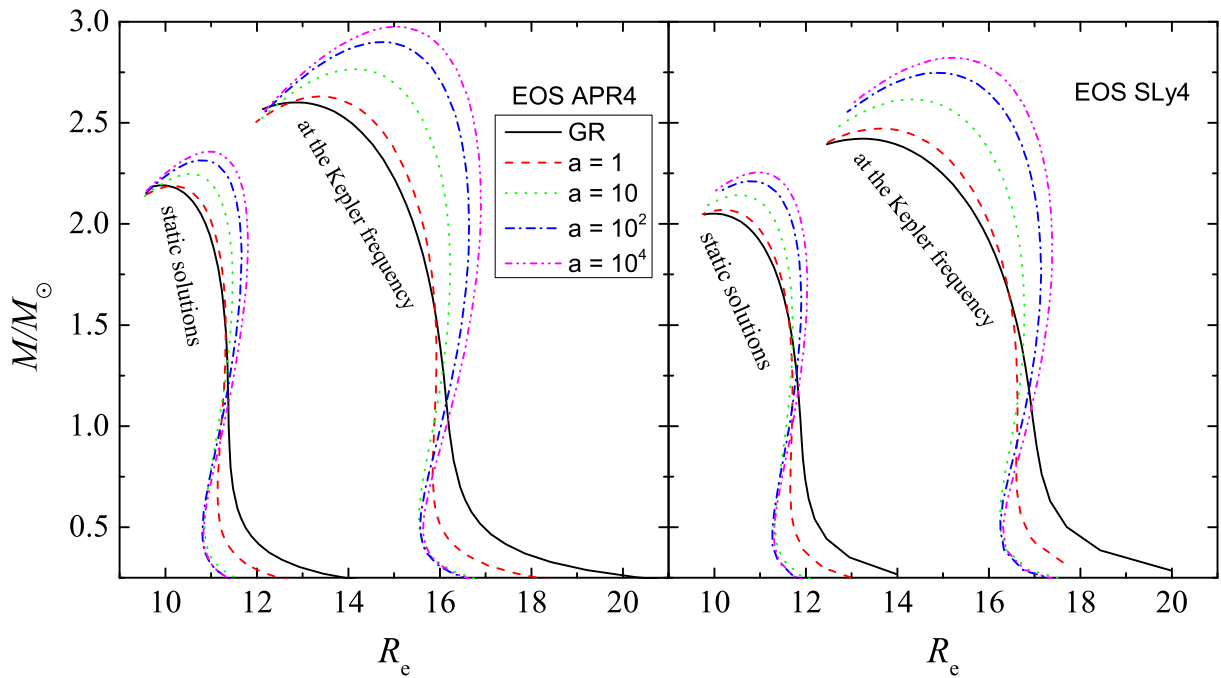}
	\caption{The mass of radius relation for EOS APR4 (left panel) and SLy4 (right panel) in the case of static neutron stars and stars rotating at the Kepler limit. Different styles and colors of the curves correspond to different values of the parameter  $a$. }
	\label{Fig:MR_AllEOS}
\end{figure}

In Fig. \ref{Fig:MR_AllEOS} the mass of radius relation is plotted for several values of the parameter $a$ in the $R^2$ theory of gravity. The limit $a \rightarrow 0$ corresponds to pure GR and the case with $a=10^4$ represents nearly the maximum possible deviation from GR (values of $a>10^4$ give results very close to the $a=10^4$ case). Two limiting sequences in terms of rotation are shown -- the nonrotating neutron star sequences and the case of stars rotating at their Kepler limit, which represents the maximum possible deviation from the static solutions. As can be seen on the figure, the rapid rotation magnifies the differences between $f(R)$ theories of gravity and GR -- the maximum masses increase considerably for large $a$ at the Kepler limit. The situation is similar to the case of scalar-tensor theories of gravity with zero potential for the scalar field which admit scalarization of the solutions \cite{Doneva2013}. In these theories considerable differences with GR are observed for rapid rotation while in the static case the scalarized solutions are very close to GR. In addition, the range of parameters where such scalarization exists is significantly broadened for fast rotation. Since the $f(R)$ theories of gravity are mathematically equivalent to a particular class of scalar-tensor theories, similar increase of the deviations compared to the static limit is expected and justified.

\begin{figure}[]
	\centering
	\includegraphics[width=0.9\textwidth]{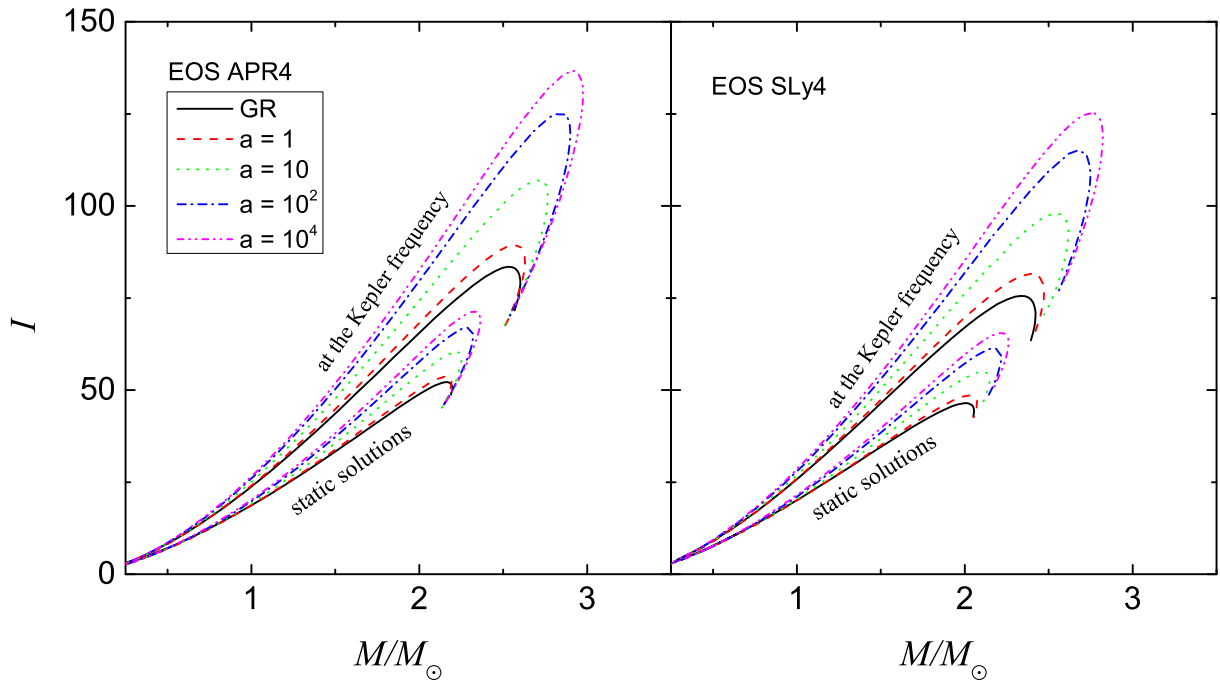}
	\caption{The moment of inertia as a function of mass for EOS APR4 (left panel) and SLy4 (right panel) in the case of static neutron stars and stars rotating at the Kepler limit. Different styles and colors of the curves correspond to different values of the parameter  $a$. }
	\label{Fig:IM_AllEOS}
\end{figure}

As the studies in the slow rotation regime showed, the $R^2$ gravity has more pronounced effect on the rotational properties of the star such as the neutron star moment of inertia \cite{Staykov2014}. The differences with GR can even exceed the EOS uncertainty for large values of the parameter $a$ that can be used in order to impose observational constraints on the theory. As expected, the rapid rotation magnifies even further these deviations, similar to the case of the $M(R)$ dependence. Indeed this can be seen on Fig. \ref{Fig:IM_AllEOS} where the moment of inertia is plotted as a function of mass for static model and for sequences of models rotating at their Kepler limit.

A quantitative estimation of the relative deviations of  the maximum mass and the maximum moment of inertia are given in Table I for the two EOS we are considering. The differences are given in percentages and for a fixed value of $a$ the quantities $\Delta M_{\rm max}$ and $\Delta I_{\rm max}$ are defined in the standard way
\begin{equation}
\Delta M_{\rm max} = \frac{M_{\rm max}^{f(R)}-M_{\rm max}^{GR}}{M_{\rm max}^{GR}}, \hskip 1cm \Delta I_{\rm max} = \frac{I_{\rm max}^{f(R)}-I_{\rm max}^{GR}}{I_{\rm max}^{GR}}.
\end{equation}
Our results show that $\Delta M_{\rm max}$ and $\Delta I_{\rm max}$ are very small for values of $a\lesssim1$. As a matter of fact if one continues decreasing the parameter $a$,  these deviations can become even negative as shown in \cite{Yazadjiev2014,Staykov2014}. Therefore the maximum mass in $R^2$ gravity can drop below the GR one for $a<1$ (for the EOS we are considering), but the deviations $\Delta M_{\rm max}$ and $\Delta I_{\rm max}$ will have very small absolute values below 1\%. Such differences can be hardly used to set any observational constrains on the $f(R)$ theories and that is why we would not pay special attention to the case when $a<1$.

\begin{table}[b] \label{Tbl:Deviations}\caption{The relative deviations of the maximum mass and the maximum moment of inertia.}
	\begin{tabular}{|r|rrrrr|}
		\multicolumn{3}{c}{APR4 Static} & \multicolumn{3}{c}{APR4 Kepler} \\
		\hline
		$a$& $\Delta M_{\rm max}$[\%] &$\Delta I_{\rm max}$[\%] &  & $\Delta M_{\rm max}$[\%] &$\Delta I_{\rm max}$[\%] \\
		\hline
		GR & 0.0 & 0.0 &  & 0.0 & 0.0 \\
		1 & 0.1 & 2.9 &  & 1.1 & 7.1 \\
		10 & 2.6 & 15.6 &  & 6.3 & 28.1 \\
		$10^2$ & 5.7 & 28.5 &  & 11.5 & 50.2 \\
		$10^4$ & 7.6 & 36.9 &  & 14.4 & 63.8 \\
		\hline
		\multicolumn{6}{c}{ } \\
		\multicolumn{3}{c}{SLy4 Static} & \multicolumn{3}{c}{SLy4 Kepler} \\
		\hline
		$a$ & $\Delta M_{\rm max}$[\%] &$\Delta I_{\rm max}$[\%] &  & $\Delta M_{\rm max}$[\%] &$\Delta I_{\rm max}$[\%] \\
		\hline
		GR & 0.0 & 0.0 &  & 0.0 & 0.0 \\
		1 & 0.9 & 4.6 &  & 2.1 & 7.8 \\
		10 & 4.5 & 18.4 &  & 8.0 & 29.6 \\
		$10^2$ & 7.8 & 32.1 &  & 13.5 & 52.4 \\
		$10^4$ & 10.0 & 41.2 &  & 16.6 & 65.6 \\
		\hline
	\end{tabular}
\end{table}

As the results show, the maximum mass increases up to 10\% in the static limit and above 16\% in the Kepler limit. As already commented in \cite{Yazadjiev2014} such deviations are non-negligible but they are comparable with the uncertainties in the nuclear matter equation of state. That is why it would be difficult to set constraints on the $f(R)$ theories using measurement of the neutron star mass and radius alone, until the EOS can be determined with smaller uncertainty. But one should keep in mind the following. Some of the proposed methods for constraining the nuclear matter EOS are related to indirect measurement of the maximum mass of rapidly rotating neutron stars. More precisely, as the results in  \cite{Bauswein2013,Takami2014} show, the observation or the lack of observation of gravitational waves signal emitted by oscillating post-merger neutron stars (a few milliseconds after the merger)  can be used to determine with a good accuracy the upper limit of the mass of stable rapidly rotating neutron stars.  But exactly this is the regime where the $f(R)$ gravity will give the most significant deviations. That is why all strategies for determining the nuclear matter EOS should be used with caution since modifications of the gravitational theory can produces similar (both qualitatively and quantitatively) effects, especially in the rapidly rotating regime where the deviations from GR are substantially magnified.

The differences in the neutron star moment of inertia on the other hand can be much more dramatic -- from 40\% increase of the maximum angular momentum in the static limit, to over 65\% in the case of rapid rotation. As commented in \cite{Staykov2014} such large deviations can be potentially measured by the forthcoming observations of the neutron star moment of inertia \cite{Lattimer2005,Kramer2009} that can lead to a direct test of the $R^2$ gravity. The gravitational wave spectrum of rotating neutron stars would be also altered considerably and such a study is underway. In particular the criteria for the development of the secular Chandrasekhar-Friedman-Schutz instability, what sets in due to the emission of gravitational radiation, can be altered significantly that would have a direct effect on the predictions for the detectability of the signal. The validity of various phenomenological EOS independent relations, connecting the neutron star mass, radius, moment of inertia, etc. \cite{Stergioulas03,Friedman2013}, including the I-Love-Q relations \cite{Yagi2013,Yagi2013a,Doneva2014,Pappas2014,Chakrabarti2014}, has to be also checked and such a study in underway.

\section{Conclusion}
In the present paper we have studied rapidly rotating neutron stars in $R^2$ theories of gravity by threating the problem non-perturbatively and self-consistently. For this purpose we have used the well known fact that $f(R)$ theories are mathematically equivalent to a particular class of scalar-tensor theories with nonzero potential for the scalar field. The numerical solutions are obtained using an extended version of the {\tt RNS} code that can deal with such scalar-tensor theories. The equilibrium properties of the solutions are studied and a detailed comparison with the static case is performed.

The results show that similar to the neutron stars in scalar-tensor theories, the rotation can magnify significantly the deviations between the $f(R)$ neutron stars and the corresponding solutions in pure GR. This effect is better pronounced for the rotational properties of the star, such as the moment of inertia and the angular momentum. The differences can be comparable or larger to the uncertainties in the nuclear matter EOS. The maximum mass of neutron stars in $R^2$ gravity can increase more than 16\% in the case of rapid rotation, while the increase of the maximum moment of inertia reaches 65\%. Clearly such large differences can have a non-negligible influence on various astrophysical phenomena, such as the observed quasiperiodic oscillations, the neutron star oscillation spectrum and the associated gravitational wave emission, etc. The deviations are in general larger for larger values of the parameter $a$ in the $R^2$ gravity, that can be used to impose an upper limit on this parameter. The various EOS independent relations between the neutron star properties might be also modified, carrying a signature from the modified theory of gravity. Such a study is underway.

Once the high density equation of state is better know, the results in the present paper can se used to set tight constraints on the $f(R)$ theories. But since rapid rotation can be observed for various astrophysical objects, such as the millisecond pulsars and the supramassive neutron stars formed after a merger, various strategies for extracting the neutron star EOS should be applied with caution. The reason is that modified theories of gravity can lead to competitive effects compared to a change in the EOS, especially in the rapidly rotating case.

\acknowledgments{D. D. would like to thank the Alexander von
Humboldt Foundation for support.  K.K. and S.Y. would like to thank
the Research Group Linkage Programme of the Alexander von Humboldt
Foundation for the support and S.Y. would like to thank the
Institute for Theoretical Astrophysics Tuebingen for its kind
hospitality. The networking support by the COST
Action MP1304 is gratefully acknowledged. }

\appendix

\section{Dimensionally reduced field equations}

In this appendix we give the Einstein frame dimensionally reduced stationary and axisymmetric field equations describing the structure of the rapidly rotating
neutron stars in $f(R)$ theories. The Einstein frame fluid energy density, pressure  and 4-velocity  will be denoted by $\varepsilon_{*}$,  $p_{*}$ and $u_{*}^{\mu}$. The  same quantities in the Jordan frame will be denoted by  $\varepsilon$, $p$ and $u^{\mu}$. The relations between the quantities in both frames are
explicitly given by $\varepsilon_{*}= A^4(\varphi)\varepsilon$, $p_{*}=A^4(\varphi) p$ and  $u_{*}^{\mu}=A(\varphi)u^{\mu}$.

In writing the dimensionally reduced equations it is convenient to use the proper velocity $v$ of the fluid given by

\begin{equation}
v = (\Omega - \omega) r \sin \theta e^{-\sigma},
\end{equation}
where $\Omega$ is the fluid angular velocity defined by
$\Omega=\frac{u^{\phi}}{u^{t}}$. In the present paper we consider only uniformly rotating neutron stars with constant angular velocity.   $\Omega$ and $v$ are the same in
both Einstein and Jordan frame \cite{Doneva2013}.  The fluid four
velocity in the Einstein frame then is
\begin{equation}
u^\mu_{*} = \frac{e^{-(\sigma + \gamma)/2}}{\sqrt{1-v^2}}
[1,0,0,\Omega].
\end{equation}

From a numerical point of view it is more convenient to use the
angular coordinate $\mu=\cos\theta$ instead  of $\theta$.
In terms of this coordinate, the dimensionally reduced field
equations for the metric functions $\gamma$, $\sigma$ and $\omega$
are the following

\begin{eqnarray}
\left(\Delta + \frac{1}{r} \partial_{r}  -
\frac{\mu}{r^2}\partial_{\mu}\right)\left(\gamma
e^{\gamma/2}\right)&=&e^{\gamma/2}\left\{(16\pi p_{*} -
V(\varphi))e^{2\alpha}  + \right. \notag \\
&& \notag \\
&&\left.+\frac{\gamma}{2}\left[(16\pi p_{*} -V(\varphi))e^{2\alpha} -
\frac{1}{2}(\partial_{r}\gamma)^2 - \frac{1}{2} \frac{1-\mu^2}{r^2}
(\partial_{\mu}\gamma)^2 \right]\right\},  \nonumber \\\label{eq:DiffEq_gamma}
\end{eqnarray}

\begin{eqnarray}
\Delta(\sigma e^{\gamma/2}) &=& e^{\gamma/2}\left\{8\pi (\varepsilon_{*}
+ p_{*})e^{2\alpha} \frac{1+ \upsilon^2}{1 - \upsilon^2}   + r^2
(1-\mu^2)e^{-2\sigma}\left[(\partial_{r}\omega)^2  +
\frac{1-\mu^2}{r^2}(\partial_{\mu}\omega)^2\right] +
\frac{1}{r}\partial_{r}\gamma - \frac{\mu}{r^2}\partial_{\mu}\gamma
 \right. \nonumber  \\ && \nonumber \\
 &&\left.  + \frac{\sigma}{2}\left[(16\pi p_{*} -V(\varphi))e^{2\alpha} - \frac{1}{r}\partial_{r}\gamma + \frac{\mu}{r^2}\partial_{\mu}\gamma -
 \frac{1}{2}(\partial_{r}\gamma)^2  - \frac{1}{2}\frac{1-\mu^2}{r^2} (\partial_{\mu}\gamma)^2 \right] \right\}, \label{eq:DiffEq_sigma}
\end{eqnarray}

\begin{eqnarray}
\left(\Delta  +  \frac{2}{r} \partial_{r}  -
\frac{2\mu}{r^2}\partial_{\mu}\right)\left(\omega e^{\gamma/2
-\sigma}\right) &=& e^{\gamma/2 - \sigma} \left\{- 16\pi
\frac{(\varepsilon_{*} + p_{*})(\Omega - \omega)}{1- \upsilon^2}e^{2\alpha}
+ \right. \nonumber \\
&&\omega\left[-\frac{1}{r}\partial_{r} (\frac{1}{2}\gamma + 2\sigma)
+ \frac{\mu}{r^2}\partial_{\mu}(\frac{1}{2}\gamma + 2\sigma) -
\frac{1}{4}(\partial_{r}\gamma)^2 -  \frac{1}{4}\frac{1-\mu^2}{r^2}
(\partial_{\mu}\gamma)^2  + \right. \nonumber \\
&& + (\partial_{r}\sigma)^2 + \frac{1-\mu^2}{r^2}
(\partial_{\mu}\sigma)^2     - r^2(1-\mu^2)e^{-2\sigma}
\left((\partial_{r}\omega)^2 + \frac{1-\mu^2}{r^2}
(\partial_{\mu}\omega)^2\right)  \nonumber \\
&& \left.  \left. - 8\pi \frac{\varepsilon_{*} (1+ \upsilon^2) +
2p_{*}\upsilon^2}{1- \upsilon^2} e^{2\alpha} -\frac{1}{2} V(\varphi)
e^{2\alpha}\right] \right\}. \label{eq:DiffEq_omega}
\end{eqnarray}

Here, the differential operator $\Delta$ is defined by
\begin{eqnarray}
\Delta =  \partial^2_{r} + \frac{1}{r}\partial_{r} +
\frac{1-\mu^2}{r^2}\partial^2_{\mu} - \frac{2\mu}{r^2}
\partial_{\mu}.
\end{eqnarray}

For the metric function $\alpha$ we have two first order partial
differential equations. In the numerical method in order to
determine $\alpha$ we need only one of them, namely the following one

\begin{eqnarray}
&&\partial_\mu \alpha = -\frac{\partial_\mu \gamma + \partial_\mu \sigma}{2} - \left\{(1-\mu^2)(1+r\partial_r\gamma)^2 + [-\mu + (1-\mu^2)\partial_\mu \gamma]^2 \right\}^{-1} \times  \label{eq:DiffEq_alpha}\\ \notag \\
&&\left\{\frac{1}{2}\left[r\partial_r(r\partial_r\gamma) + r^2 (\partial_r \gamma)^2 - (1-\mu^2)(\partial_\mu \gamma)^2 - \partial_\mu[(1-\mu^2)\partial_\mu \gamma] + \mu\partial_\mu \gamma\right] \times [-\mu + (1-\mu^2)\partial_\mu \gamma] + \right. \notag \\ \notag \\
&& +\frac{1}{4}[-\mu + (1-\mu^2)\partial_\mu \gamma] \times \left[r^2(\partial_r \gamma + \partial_r \sigma)^2 - (1-\mu^2)(\partial_\mu \gamma + \partial_\mu \sigma)^2 + 4r^2(\partial \varphi)^2 - 4(1-\mu^2)(\partial_\mu \varphi)^2\right] + \notag \\ \notag \\
&& + \mu r \partial_r\gamma [1+r\partial_r\gamma] - (1-\mu^2)r(1+r\partial_r\gamma)\left[\partial_\mu\partial_r \gamma + \partial_\mu\gamma\partial_r\gamma + \frac{1}{2}(\partial_\mu \gamma + \partial_\mu \sigma)(\partial_r \gamma + \partial_r \sigma) + 2 \partial_\mu\varphi\partial_r \varphi \right] +\notag \\ \notag \\
&&+\frac{1}{4}(1-\mu^2)e^{-2\sigma}\left[-[-\mu+(1-\mu^2)\partial_\mu
\gamma][r^4(\partial_r \omega)^2 - r^2(1-\mu^2)(\partial_\mu
\omega)^2] + \right. \notag \\ \notag \\
&&\left.\left.+2(1-\mu^2)r^3\partial_\mu\omega\partial_r\omega(1+r\partial_r\gamma)\right]\right\}.
\notag
\end{eqnarray}

The dimensionally reduced equation for the scalar field $\varphi$ is

\begin{eqnarray}
\Delta \varphi= - \partial_{r}\gamma\partial_{r}\varphi -
\frac{1-\mu^2}{r^2} \partial_{\mu}\gamma\partial_{\mu}\varphi +
\left[-\frac{4\pi}{\sqrt{3}}(\varepsilon_{*} - 3p_{*}) +
\frac{1}{4}\frac{dV(\varphi)}{d\varphi}\right]e^{2\alpha}.
\label{eq:DiffEq_phi}
\end{eqnarray}

The above equations have to be supplemented with the  equation
for  fluid equilibrium

\begin{eqnarray}
\frac{\partial_i{p}}{\varepsilon + p} -
\left[\partial_i(\ln \, u^t_{*})  +
 \frac{1}{\sqrt{3}}\partial_i \varphi\right]=0 \label{eq:Hydrostatic_Equil}.
\end{eqnarray}

\bibliography{references}

\end{document}